\documentclass[10pt,preprint]{aastex}

\shorttitle{Torus/Jets of PSR B1706$-$44}
\shortauthors{}

\begin{document}

\title{The Complex Wind Torus and Jets of PSR B1706$-$44}

\author{Roger W. Romani$^1$, C.-Y. Ng$^1$, Richard Dodson$^2$ \& 
Walter Brisken$^3$}
\affil{$^1$Department of Physics, Stanford University, Stanford, CA 94305}
\affil{$^2$The Institute of Space and Astronautical Science,
                 3-1-1 Yoshinodai,
                 Sagamihara, Kanagawa 229-8510, Japan}
\affil{$^3$National Radio Astronomy Observatory, P.O. Box O, Socorro, NM 87801}
\email{rwr@astro.stanford.edu, ncy@astro.stanford.edu, rdodson@vsop.isas.jaxa.jp, wbrisken@aoc.nrao.edu}

\begin{abstract}

We report on {\it Chandra} ACIS imaging of the pulsar wind nebula (PWN)
of the young Vela-like PSR B1706$-$44, which shows the now common pattern
of an equatorial wind and polar jets. The structure is particularly rich,
showing a relativistically boosted termination shock, jets with strong
confinement, a surrounding radio/X-ray PWN and evidence for a quasi-static
`bubble nebula'. The structures trace the pulsar spin geometry and
illuminate its possible relation to SNR G343.1$-$2.3. We also obtain
improved estimates of the pulsar flux and nebular spectrum, constraining
the system age and energetics.
\end{abstract}


\section{Introduction}

	PSR B1706$-$44, discovered by \citet{jet92} is among the most
interesting pulsars for study at high energies. It is one of a handful of
pulsars detected by EGRET in GeV $\gamma$-rays. It is quite similar to the
Vela pulsar with a characteristic age $\tau_c = {P/(2{\dot P})} = 
1.7 \times 10^4$\,yr and a spindown luminosity of ${\dot E} \approx 4 
\times 10^{36}$\, erg/s, but is $\sim 10\times$ more distant at $d = 3d_3$\,kpc. 
Early {\it Chandra} HRC/ACIS data provided a first detection of X-ray
pulsations and showed a compact $\sim 10^{\prime\prime}$ surrounding pulsar 
wind nebula (PWN) \citep{ghd02,dg02}. More recent {\it XMM-Newton}
spectroscopy \citep{met04} has provided improved measurements of the
X-ray spectrum and pulsations.  Early claims that the PWN is detected in TeV
$\gamma$-rays \citep{ket95,chet97} have not been supported by recent HESS
observations \citep{hess05}.

	PSR B1706$-$44 is superposed on a radio-bright spur of the 
supernova remnant G343.1$-$2.3, which has a similar, albeit unreliable,
$\Sigma-D$ distance of $\sim 3$\,kpc \citep{mo93}. The pulsar DM gives a distance
of $2.3\pm0.3$\,kpc in the \citet{cl02} model. 
\citet{dg02} have argued for an association. 
In particular, they found a faint southern extension of the SNR, which would
place the pulsar within the full SNR boundary. They also noted an approximately
N-S elongation of the X-ray PWN, pointing roughly back to the SNR center and
argued that this would represent a trailed nebula. The required velocity
for travel from the approximate geometric center of the SNR, about $12^\prime$
away, was $\sim 1000 d_3/\tau_4$\,km/s  where $\tau_4$ is the pulsar
age in units of $10^4$\,yr. There are, however, some challenges to 
this SNR association. \citet{koet95} in an HI absorption study of the pulsar
found velocity components setting lower and upper bounds for the distance
of $d_{min} = 2.4\pm0.6$\,kpc and $d_{max}=3.2\pm0.4$\,kpc. However a prominent
HI emission feature seen in the bright limb of G343.1$-$2.3 at $-32$km/s is not seen 
by Koribalski et al. in the absorption spectrum of the
pulsar, suggesting that it lies in front of the SNR. Also scintillation
studies \citep{net96,jnk98} suggest a low transverse velocity for the pulsar,
$v \le 89$\,km/s. This estimate has been supported by more recent scintillation 
measurements (Johnston, priv. comm). Thus the distances of the pulsar and SNR
are still fairly uncertain. We adopt here a generic distance of 3\,kpc in
the discussion that follows, but carry through the scaling to show the distance
dependence.

\begin{figure}[ht!]
\epsscale{1.0}
\plotone{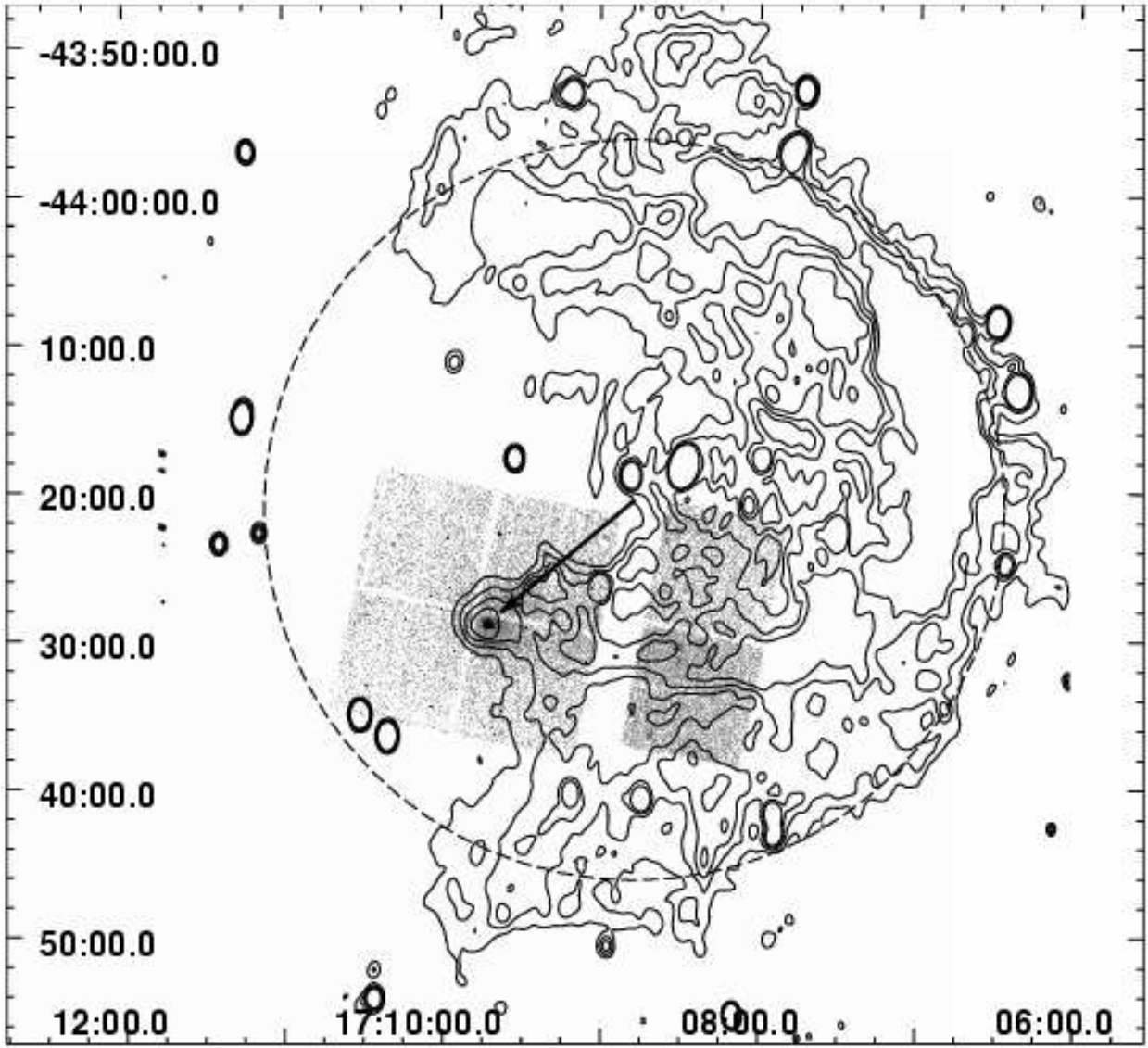}
\caption{Greyscale image of our new ACIS-I pointing of PSR B1706$-$44. The
contours (at 8, 10, 12 and 14 mJy/beam) show the shell of G343.1$-$2.3 from a 
19 pointing 1384\,MHz ATCA mosaic \citep{dg02}. The radio map has a resolution
of $70^{\prime\prime} \times 47^{\prime\prime}$ and an rms final map noise of
0.6mJy.  The X-ray PWN lies on a spur of radio emission.
An approximate boundary of the full SNR (25$^\prime$ radius) and
an arrow for the inferred PSR motion, assuming birth at the SNR center, 
are shown.
}
\end{figure}

	\citet{nr04} re-examined the AO1 50\,ks HRC and $\sim 15$\,ks
ACIS-S exposures, and found that the compact PWN could be well modeled as
an equatorial torus + polar jets, rather similar to the structures seen
around the young Crab and Vela pulsars. They found that the brightest arc
of PWN emission lay {\it behind} the pulsar's inferred motion from the SNR
center, while a polar jet extended well in front of the pulsar position.
This is difficult to reconcile with a bow shock/trail interpretation. On
the other hand, the PWN symmetry axis did indeed point back to the SNR
center, suggesting that a more careful evaluation of the connection was in 
order. This is particularly interesting, since comparison of the PWN
symmetry (= pulsar spin) and proper motion axes can constrain the
origin of pulsar birth  kicks \citep{sp98,lcc01,r04}.

	We have obtained a deeper 100\,ks ACIS-I exposure of PSR B1706$-$44
and surroundings. The X-ray exposure coverage is compared to the overall geometry
of G343.1$-$2.3 in Figure 1.
Together with new ATCA radio continuum imaging
we are able to study the rich structure in this PWN and further constrain
its connections with the SNR.
	
\bigskip

\section{Observations and Data Analysis}

	PSR B1706$-$44 was observed with the {\it Chandra} ACIS-I array 
(4 ACIS-I chips along with the S3 and S4 chips) on February 1-2, 2004 with 
standard imaging (3.2s TE) exposures. The CCD array was operated in `very faint' 
(VF) mode, allowing improved rejection of particle backgrounds. The total 
live-time was 98.8\,ks and no episodes of strong background flaring were observed. 
Hence all data are included in our analysis.
The pulsar was positioned near the standard aimpoint of the I3 chip and all
observing conditions were normal. We have also compared our new exposure
with the archival (February 3, 2001) 14.3\,ks ACIS-S3 exposure (obsID 0757). As
usual the backside illuminated S3 chip suffered more from particle background 
and after cutting out periods of background flares, 11\,ks of clean exposure remained.
All analysis was performed using CIAO 3.2 and CALDB 3.0.0, including automatic correction
for the ACIS QE degradation. These data were nearly free of pile-up; the maximum
pixel counts at the pulsar position indicate only $2.5$\% pile-up while
the best fit model for the point source has an expected pile up fraction of
$\sim 3.5$\%. For sources with low pile-up we can maximize the spatial resolution 
of the ACIS image by removing the standard pixel randomization and applying an
algorithm correcting the position of split pixel events \citep{mo01}. This decreases 
the on-axis PSF width in our data set by $\ga10$\%. These data are compared with
radio observations of the PWN.

\subsection{Radio Imaging and Astrometry}

	Data for the radio maps shown here were collected at the Australia 
Telescope Compact Array (ATCA) in Narrabri (latitude -30.3$^\circ$) \citep{atca92}. 
For the 1.4\,GHz map in Figures 1 and 2a, the data acquisition and
analysis are described in \citet{dg02}. For the image contours in Figure 3a,
the data first presented in \citet{dg02} were re-imaged including the 6km baselines
and uniform weighting to highlight the high resolution features. The restoring beam 
size is $9\farcs0 \times 7\farcs8$. To show the nebular structure, an 11mJy point 
source PSF has been subtracted at the position of the pulsar. Two maxima appear
flanking the pulsar position. These are unlikely to be artifacts due to pulsar
variability, as diffractive scintillation for this pulsar is particularly weak
\citep{jnk98}. Since the data were collected in five sessions, spread over more
than a year, it is in principal possible for slow refractive scintillation to
change the pulsar flux between epochs and distort its PSF. However, each epoch
used $\sim$12h of integration, so any residual epoch PSF should be close to 
circularly symmetric, in contrast to the structure near the pulsar which is
clearly bipolar. Further observations, with pulsar binning, have been requested 
to confirm this result.

For the 4.8\,GHz map in Figure 2b, observations
were made at 4.8- and 8.6\,GHz with the array in the standard
configurations 0.75A, 1.5A and 6A on 06 Jan, 16 Feb and 11 Apr 2002. The
maximum and minimum baselines
for the 4.8-GHz data were 1 and 100 k$\lambda$ (angular resolutions of
$3.4^{'}$ to $2\farcs1$) for a total of 26 hours observation. 
In all cases we
observed the two frequencies with bandwidths of 128 MHz. We used the
ATNF correlator mode that divides each integration's data into separate
phase bins spanning the pulsar period.  This firstly allowed the
strongly pulsed point source flux to be excluded from the image and secondly
allowed us to self-calibrate using the relatively strong point source
flux from the pulsar. After data editing and
calibrating we inverted the image with a {\em uv}-taper of $20^{''}$
and deconvolved it with the full polarization maximum entropy task
{\small PMOSMEM} in {\small MIRIAD}. 

The most important test of the SNR association would, of course, be a 
direct astrometric proper motion. With a 1.4\,GHz flux of $\sim$11\,mJy,
PSR B1706$-$44 is relatively bright.  As such it is suitable for 
phase referenced VLBI astrometry, if an in-beam reference could be 
found. Unfortunately searches for phase references adequate for 
Australian Long Baseline Array (LBA) and US 
VLBA experiments have not detected comparison sources with compact 
fluxes greater than $\sim 1$mJy. Attempts were made at external
phase reference VLBA astrometry. However at 1.4\,GHz, the nearest known 
reference source (2.5$^\circ$ away) was scatter broadened to 
$\sim 50$\,mas. With the strong ionosphere at such low elevation,
the next nearest known source (10$^\circ$ away) is too distant for
effective calibration. Since the pulsar spectrum is steep, an attempt
at VLBA astrometry at 5\,GHz was also unsuccessful; at this low elevation the system
temperature was $4-5\times$ nominal and only six VLBA antennae could be used,
reducing the sensitivity to $\sim 15\%$ of nominal.
So unfortunately we have only
tied-array astrometry at present. Even if the pulsar does travel from the
geometric center of G343.1$-$2.3, the expected proper motion is only $\sim 40$\,mas/yr;
the existing time base of VLA/ATCA imaging does not yet
allow a serious constraint on this motion.  We must conclude that a 
direct proper motion measurement awaits substantially increased (SKA or EVLA)
capabilities and a long-duration, large base-line experiment.
%
\subsection{X-ray Spatial Analysis}

	To show the diffuse emission surrounding PSR B1706$-$44 we plot (Fig. 2a)
a 1-7keV image with point sources removed (except the pulsar). These data are 
exposure corrected to minimize the chip gaps and heavily smoothed on
a $20^{\prime\prime}$ scale. The diffuse emission is an edge-brightened, radius
 $\sim 110^{\prime\prime}$ cavity surrounding the pulsar with a faint extension 
to the west. Contours of the 1.38\,GHz radio map show good
correlation with the radio emission in the bar crossing G343.1$-$2.3 (Fig. 1).
We will refer to this structure as the `nebula'. 

\begin{figure}[ht!]
\includegraphics[scale=0.45]{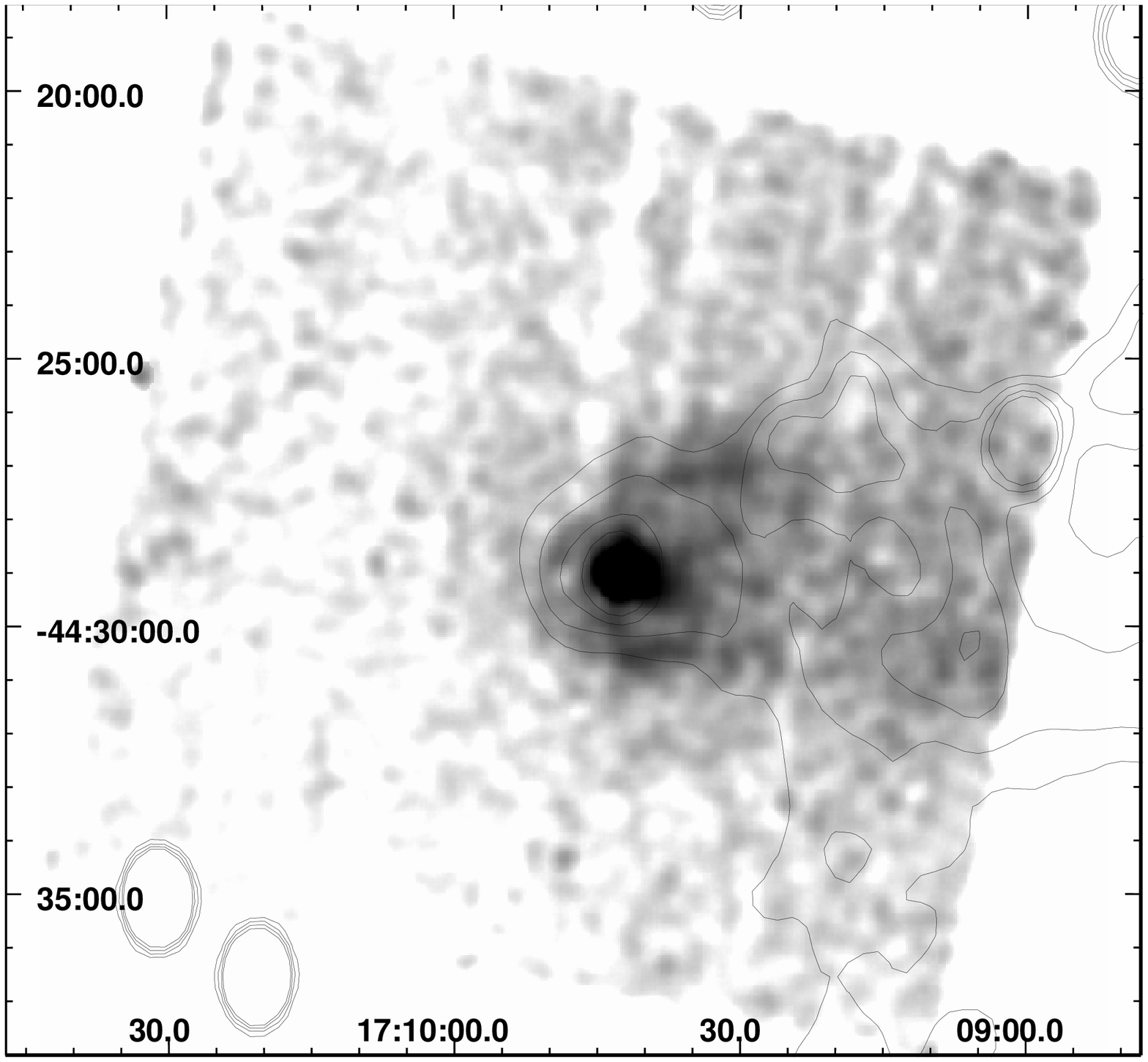}
\includegraphics[scale=0.45,clip=true]{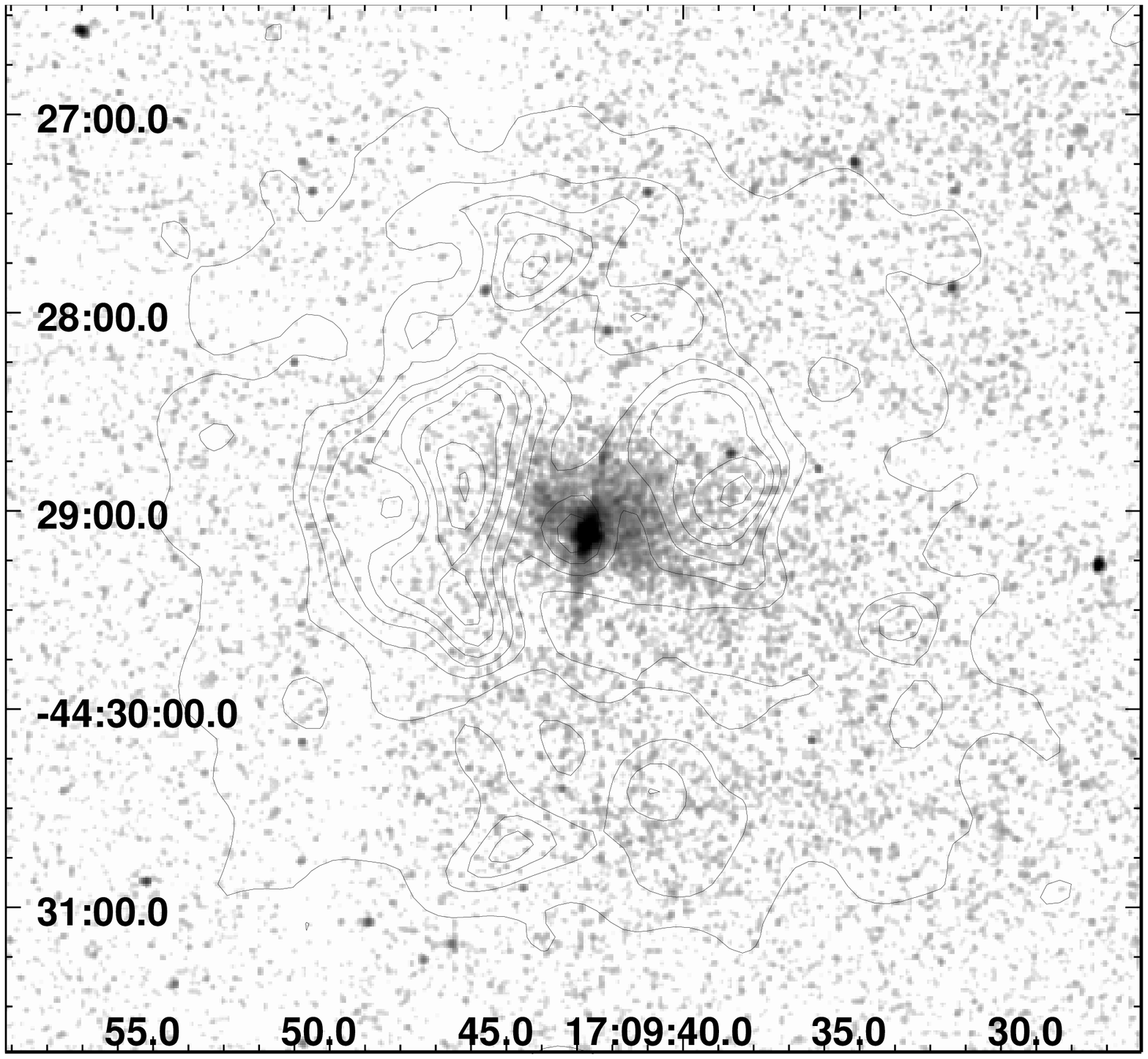}
\caption{Left: ACIS-I 1-7\,keV image with point sources (other than the pulsar)
removed, exposure correction and $20^{\prime\prime}$ Gaussian kernel smoothing.  
Contours are from the 1.38\,GHz radio map of Figure 1. 
Right: $1\farcs5$ Gaussian smoothed image of the PWN with an overlay of the
core of the radio nebula from a 4.8\,GHz ATCA image (contours at 0.4, 0.8, 1.0,
1.1,...1.6  mJy/beam); the
resolution is $20^{\prime\prime}$ and the image RMS is 0.2mJy/beam.
}
\end{figure}

\bigskip

	Moving in toward smaller scale, in 2b we show a 1-7keV image, smoothed
with a $1\farcs5$ Gaussian. Point sources have not been removed. This
shows that the cross structure
fit by \citet{nr04} extends across $\sim 1^\prime$. Narrow X-ray jets, which we
refer to here as the `outer jet' (extending south) and `outer counter jet'
(extending north) start $\sim 10^{\prime\prime}$ from the pulsar and continue
to $\sim 30^{\prime\prime}$. Bracketing these is faint diffuse X-ray emission 
which we will call the `equatorial PWN'. For comparison we draw contours of
a 4.8\,GHz ATCA image with $21^{\prime\prime} \times 18^{\prime\prime}$ 
restoring beam. These observations have the pulsar `gated out' and show that
the radio PWN has a hollow center bracketing the equatorial PWN. Diffuse radio peaks
are, in fact, seen just east and west of this X-ray structure.

	Finally, we show in Figure 3 a lightly smoothed image
of the central region of the PWN, stretched to bring out the faint outer jets.
The contours are drawn from a 1.38\,GHz ATCA image, where the
6-km baselines have been weighted to produce a $9\farcs0 \times 7\farcs8$ restoring
beam. A point source PSF has been subtracted at the pulsar position. Two
local radio maxima with peak fluxes $\sim 2$mJy and $\sim 2.5$mJy bracket the `torus'
structure. The radio then shows a sub-luminous zone surrounding the
`equatorial PWN'; beyond $\sim 30^{\prime\prime}$ the radio brightens again,
as in figure 2b. No emission appears along the `outer jets'. Indeed there appear
to be evacuated channels in the radio emission, but improved S/N and resolution
are needed to probe this sub-mJy structure.
The frames on the right show the innermost region of the PWN with the best-fit
torus + inner jet model (\S2.3).

\begin{figure}[b!]
\includegraphics[scale=0.50,origin=200]{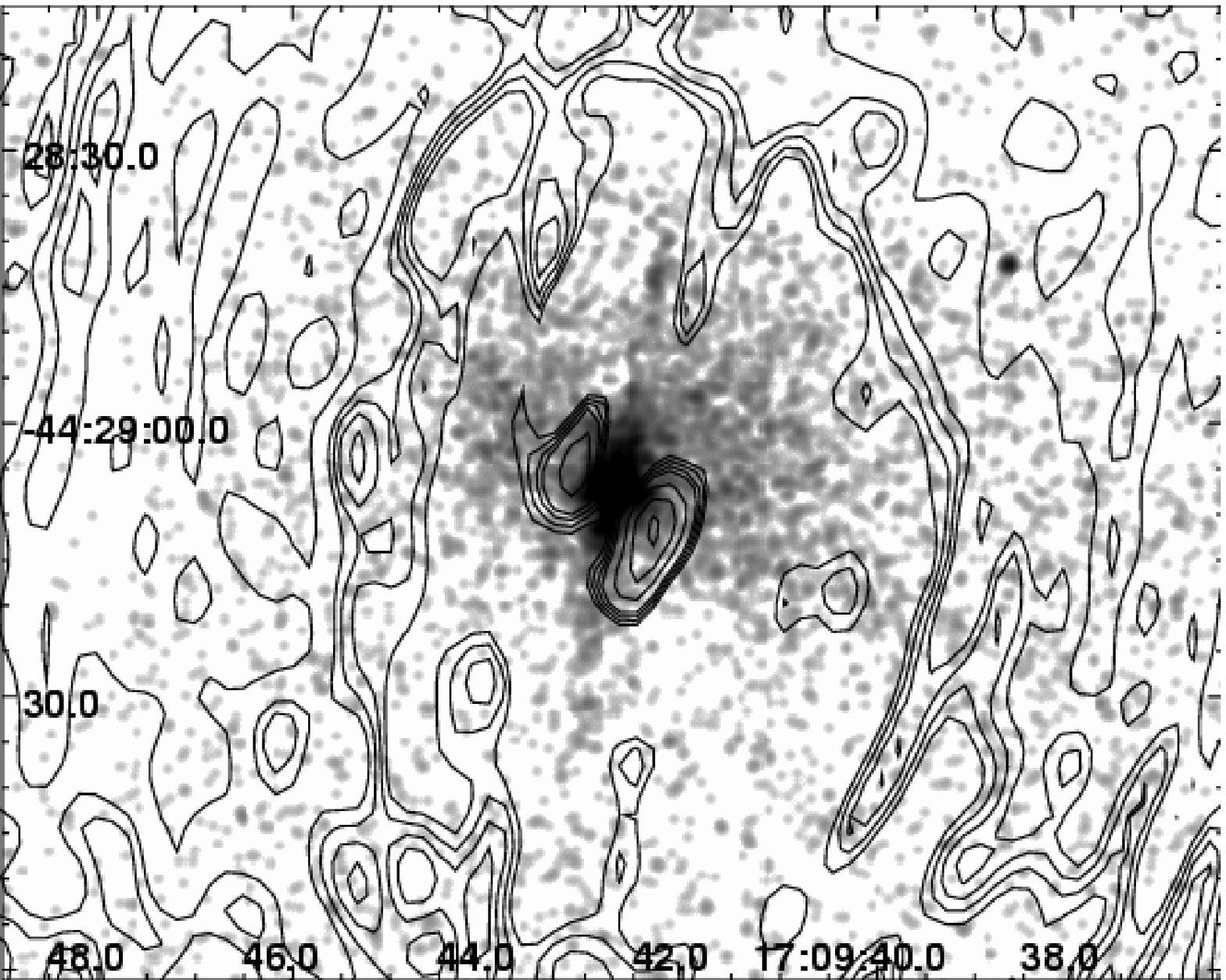}
\includegraphics[scale=0.28,clip=true]{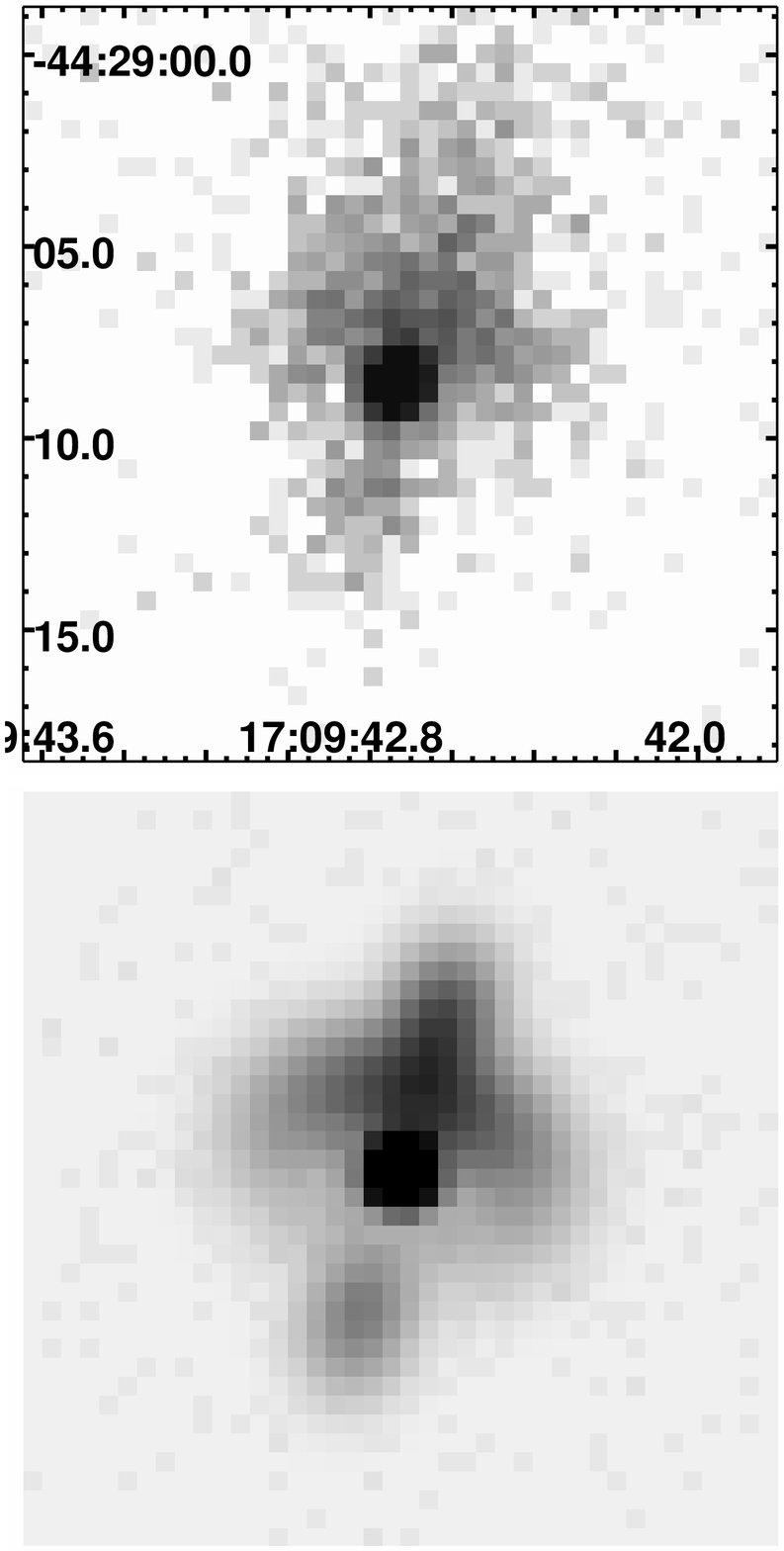}
\includegraphics[scale=0.34,clip=true]{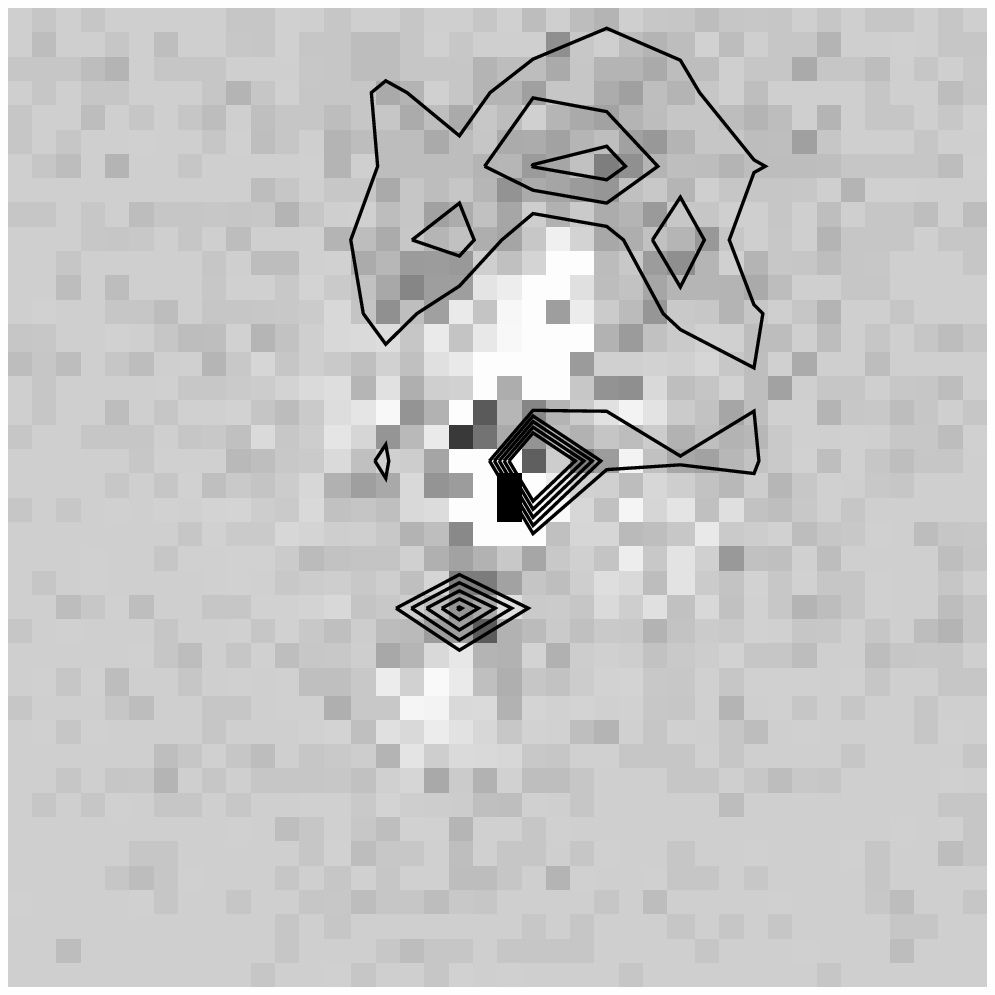}
\caption{Left: Grey scale image ($1\farcs5$ Gaussian smoothing) of the PWN,
stretched to show the outer jets and the equatorial PWN. The contours 
(0.3, 0.4, 0.5, 0.6, 0.9, 1.2, 1.5\,mJy/beam) are from the 1.38\,GHz ATCA image, 
which has a map rms of 0.2 mJy/beam and a resolution of $9\farcs0 \times 7\farcs8$.
The pulsar point source has been subtracted.  Two peaks of radio emission bracket 
the torus. These lie in a cavity which surrounds the equatorial PWN and jets.
Middle: grey scale image of the inner PWN (above) and of the best-fit point 
source+torus+(inner)jets model (Table 1) to the same scale(below). 
At the far right, the residual image is shown, with contours indicating the 
excess counts above (North of) the torus.
}
\end{figure}

\bigskip

	The overall geometry of the PWN is strongly reminiscent of that 
surrounding the Vela pulsar. In particular \citet{pet03} have described a series
of ACIS images of the Vela nebula which show a torus-like structure, an inner jet and
counter jet and a faint narrow outer jet system.  This imaging sequence showed
that the Vela outer jet, which is patchy and strongly bent, varies dramatically on
timescales of days to weeks. Apparent motion of blobs within the jets suggests mildly
relativistic bulk velocities and strong instabilities. For PSR B1706$-$44 our
single sensitive image does not let us comment on variability. However we
argue that the relatively straight and narrow jets, $\sim 3\times$ longer than 
those of Vela, and symmetric PWN structure are a consequence of a static
uniform external environment and a low pulsar velocity. At 1.4-8.5\,GHz
\citet{doet03}
have found that the Vela PWN has two bright patches bracketing the X-ray
torus and jets, in a structure quite similar to that in Figure 2b. Polarization imaging
of the Vela radio structure suggests that these two patches represent the limbs of
a toroidal B field structure. This implies that the rotation axis
controls the PWN symmetry to large radii.

\subsection{Nebula Structure Fits}
Following \citet{nr04} we have fitted our new ACIS image to a point source PSF,
Doppler boosted equatorial torus, polar jets and uniform background. The 
fitting minimizes residuals using a Poisson-based likelihood function. Monte 
Carlo simulations of Poisson realizations of the best-fit model are 
in turn re-fitted to generate statistical errors and their co-variance matrices. 
Table 1 contains the best-fit values. The torus radius and axis inclination and
position angles are $r$, $\zeta$ and $\Psi$, respectively. See \citet{nr04}
for the definition of the other parameters and the details of the fitting
technique. In Table 1 the inner jet/counter-jet are constrained to lie along 
the torus axis in the fits.

\begin{table}[!ht]
\caption{Torus Fit Parameters with 1$\sigma$ Statistical Errors}
\medskip
\tabletypesize=\small
\begin{tabular}{ccccccccccc}
$\Psi$ (deg) & $\zeta$ (deg) & r (arcsec) & $\delta$ & $\beta$ &  Point Source$^\dagger$ & Torus$\dagger$ & Jet$\dagger$ & Counter-jet$\dagger$ \\ \hline \hline
$163.6\pm0.7$ & $53.3_{-1.4}^{+1.6}$ & $3.3_{-0.06}^{+0.08}$ & $1.0^*$ & $0.
70\pm0.01$ & $2897_{-52}^{+65}$ & $1221_{-41}^{+55}$ & $185_{-21}^{+8}$
 & $325_{-54}^{+43}$ 
\end{tabular}
\leftline{ * -- held fixed \qquad \qquad $\dagger$ counts }
\label{torus}
\end{table}

%
%

	In addition to the statistical errors, there are certainly systematic
errors, in particular induced by unmodeled PWN components. For example,
it is clear that there are counts in excess of the torus+jet model in a
cap surrounding the inner counter-jet. Interestingly similar structure is seen in 
the Crab PWN. We have made an attempt to constrain the systematic biases by modifying the
fitting model. For example, allowing the (inner) jet and counter-jet to have a free
position and amplitude shifts the best-fit position angle to
$\Psi =165\pm 0.5^\circ$ and the inclination to $\zeta = 56.7\pm1.0^\circ$. 
We therefore infer systematic errors about $3\times$ larger than our rather 
small statistical errors.

	We have also measured the outer jet/counter-jet system. Minimizing
the residual to a 1-D line passing through the pulsar, the two jets together 
lie at $\Psi_{\rm outer} =169.4\pm0.15^\circ$. If the jets are fitted 
separately, we obtain $\Psi_\mathrm{outer}=168.4\pm0.2^\circ$ and 
$170.9\pm0.2^\circ$ for the outer jet and counter-jet, respectively.
Thus the two jets are mis-aligned at the 
$\sim 8\sigma$ level. A fit to the count distribution about the best fit axis
shows that the narrow outer counter-jet has a Gaussian FWHM across the jet of 
$2\farcs3 \pm0\farcs2$. The outer jet appears broader at the base with an 
initial width of $4\farcs9 \pm0\farcs5$, continuing at
FWHM=$2\farcs7 \pm0\farcs3$ for its outer half. These estimates have been
corrected for the telescope PSF, which is quite uniform this close to the 
aimpoint.

	It is important to note that at the observation roll
angle, the read-out direction lies at $\Psi=168.7^\circ$! Due consideration,
however, shows that the jet structure cannot be produced by the read-out 
trail. First the jets 
cover only $\sim 1^\prime$; the read-out excess should cover the full I3 
chip. Second the pulsar provides only $\sim 2900$ $1-7$\,keV counts. The read-out trail
(out-of-time) image of this source should contribute only 36 counts over
the full $8.3^\prime$ strip across I3 and $\sim 2.5$ counts in the `jet regions';
the outer jet and counter-jet have $92$ and $93$ $1-7$\,keV counts, 
respectively. Finally the outer jets are much
harder than the soft X-ray emission from the pulsar; indeed with a mean
detected photon energy $\sim 2$\,keV these are the hardest extended features 
in the image.

	The over-all system, showing an asymmetric torus, broad inner jets
and narrow outer jets is, of course, very similar to the Vela PWN as
studied with {\it Chandra} by \citet{pet03}. We will discuss the comparison
with the Vela system in \S3, highlighting the differences. We interpret these
as suggesting that the PSR B1706$-$44 PWN has developed from a low velocity 
pulsar.

\subsection{Spectral Analysis}

	For the best possible constraints on the source spectrum, we have reprocessed
both the $\sim 11$\,ks cleaned ACIS-S data set and our new $\sim 100$\,ks 
ACIS-I data set with the new time-dependent gain adjustment and CTI correction
available in CIAO 3.2. The updated RMFs should in particular improve the low
energy calibration, important for obtaining the best estimates of $N_H$.
As noted above, in these data sets the pile-up was negligible at $\sim3\%$. To 
model the aperture corrections, 10 PSFs with monochromatic energies from 
0.5 to 9.5keV were simulated using the Chandra Ray-Tracer program, ChaRT.
The enclosed energy fraction as a function of radius was fitted to
a linear function of energy and this was used to correct the ARFs used
in the spectral fit.  In extracting the pulsar spectrum,
an aperture of radius 1$^{\prime\prime}$ was used to minimize nebular contamination.
Results from the combined fits of the ACIS-I and ACIS-S pulsar data sets are listed 
in Table~1; the spectral fits are substantially
better for composite models with both thermal and power-law components. 
Spectral parameter errors are projected multi-dimensional $1\sigma$ values. 
We quote both absorbed and unabsorbed fluxes.  As is often the case with low statistics
X-ray spectra, projected (multi-dimensional) errors on the fluxes are
very large due to spectral parameter uncertainties. Thus, we follow other
authors in quoting flux errors as $1\sigma$ single parameter values.

\begin{table}[!h]
\caption{Spectral Fits to PSR B1706$-$44}
\tabletypesize=\scriptsize
\begin{tabular}{lccccccccc}
Model&N$_H$&$\Gamma$&abs.flux&unabs. flux&T$^\infty$&R$^\infty$&abs. flux&unabs. flux&$\chi^2$/dof \\
&$10^{21}$cm$^{-2}$& &$f_{0.5-8}^\dagger$&$f_{0.5-8}^\dagger$&MK&km&$f_{0.5-8}^\dagger$&$f_{0.5-8}^\dagger$& \\
\hline
PL & $3.8\pm0.3$ & $3.06_{-0.14}^{+0.15}$ & $2.01\pm0.04$ & $4.74\pm0.1$ & - &  - & - & - & 145.9/99 \\
BB & $1.0\pm0.3$ & - & - & - & $3.7\pm0.2$ & $0.4_{-0.11}^{+0.13}$& $1.42\pm0.03$ & $1.76\pm0.04$ & 288.4/99 \\
PL+BB& $4.7_{-0.7}^{+0.8}$ & $1.62\pm0.2$ & $1.70\pm0.07$ & $2.25\pm0.1$ & $2.00_{-0.16}^{+0.17}$ & $2.4_{-1.0}^{+1.6}$& $0.83\pm0.03$ & $3.85\pm0.12$ & 40.6/97 \\
 & $5.0^*$ & $1.68\pm0.17$ & $1.70\pm0.07$ & $2.33\pm0.1$ & $1.93\pm0.06$ & $2.8_{-0.70}^{+0.76}$& $0.8 2\pm0.03$ & $4.40\pm0.14$ & 40.8/98 \\
PL+Atm& $5.9_{-0.8}^{+0.9}$ & $1.67\pm0.2$ & $1.67\pm0.07$ & $2.34\pm0.1$ & $0.79_{-0.11}^{+0.13}$ & $13.1^*$& $0.85\pm0.03$ & $6.8\pm0.2$ & 40.7/97 \\
 & $5.0^*$ & $1.56\pm0.17$ & $1.67\pm0.07$ & $2.2\pm0.1$ & $0.91\pm0.05$ & $13.1^*$& $0.88\pm0.03$ & $ 4.92\pm0.15$ & 40.8/98 \\
\end{tabular}
$^\dagger$ 0.5-8\,keV fluxes in units of $10^{-13}{\rm erg/cm^2/s}$ \qquad $^*$ held fixed
\label{psrspec}
\end{table}

	To get the best constraints on the point source spectrum, Table ~\ref{psrspec} gives
fits with $N_H$ held to the value from the power law fits to the extended emission. We
have also compared our results with the {\it XMM-Newton} fitting of \citet{met04},
by fitting counts in the $20^{\prime\prime}$ aperture used in that observation.
Our parameters and fluxes for the thermal component are generally in very good 
agreement. However, since this aperture contains much of the torus and central PWN,
{\it XMM-Newton} substantially overestimates the non-thermal flux for the
point source. Their fit power law flux corresponds to $7.5 \times 10^{-13}
{\rm erg/cm^2/s}$\,(0.5-8keV, unabsorbed). In the $20^{\prime\prime}$ aperture we find 
$9.6 \times 10^{-13} {\rm erg/cm^2/s}$\,(0.5-8keV); while the small {\it Chandra}
point source aperture gives $2.3 \times 10^{-13} {\rm erg/cm^2/s}$\,(0.5-8keV) for
the power law component. We find a similar $\sim 3\times$ excess in the power-law
+ atmosphere flux for the fit to the large {\it XMM-Newton} aperture.  Conversion 
of the power law flux observed in our small point source aperture to the {\it XMM-Newton} 
band shows that the expected PN+MOS\,(0.2-10\,keV) count rate is 18\% of the
total (power law + thermal) counts in the $20^{\prime\prime}$ aperture. 
In soft (0.2-1.35\,keV) and hard (1.35-10\,keV) bands the predicted fraction 
of the counts from the power law are 12\% and 21\%, respectively. However
the light curves of \citet{met04} show that the pulse fractions are 
21\% (soft), 12\% (hard) and 11\% (total). Since the small aperture power law
produces only 12\% of the soft counts but 21\% are pulsed, there must a 
thermal pulse component. Conversely, since the power law produces 21\% of the hard
band flux, but this only has a pulse fraction of 12\%, some of the power law
counts must be unpulsed. Extrapolation of the PWN count excess above the
point source PSF in the sub-luminous zone at 2$-3^{\prime\prime}$ produces 
$\la 1$\% of the point source aperture counts. Thus the larger scale torus
emission does not contribute significantly to the point source power law
and cannot account for its unpulsed component. This suggests that part
of the magnetospheric emission is nearly isotropic or that there is a
very compact ($\la 1^{\prime\prime}$) PWN component at the pulsar position.

For the thermal component, the fit flux gives an emitting area (effective radius)
as a function of distance. Our fit to a pure blackbody gives $R_{eff} = 2.8 d_3$\,km. 
Thus for reasonable distances, this flux represents hot $T \sim 2 \times 10^6$\,K
emission from a small fraction of the stellar surface ($\sim 4.5$\% for an 
$R^\infty = 13.1$\,km star).  The light element neutron star atmosphere models,
such as the pure H $10^{12}$\,G model grid used here \citep{zet96}, have large Wien
excesses. When fit they give lower $T_{eff}$. Also, the black body departures allow
one, in principle, to fit both the surface redshift and radius. In practice,
these are typically highly degenerate in CCD-quality data. We assume here a generic
surface radius of $R_s = 10$\,km, corresponding to 
$R_\infty = R_s (1-2GM/R_sc^2)^{-1/2} = 13.1$\,km.  With $N_H$ free (giving
$5.9\pm 0.9 \times 10^{21} {\rm cm^{-2}}$) our thermal flux normalization gives a radiating
radius of $R_\infty = 27.4 d_3$km, which is difficult to reconcile with
expected neutron star radii for any $d>1.8$\,kpc. However, when $N_H$ is
fixed at the nebular value of $5\times 10^{21} {\rm cm^{-2}}$, we get
an effective radius of $R_\infty = 16.1 d_3$km, which is tolerable even
at our nominal 3\,kpc distance.

Analysis of the low 
signal-to-noise, extended flux depends critically on the background subtraction.
Given the limited statistics, only
simple absorbed power-law fits were attempted for all non-thermal sources.
The results are listed in Table~\ref{pwnspec}.  For consistency, 
all fits are to the 0.5-8\,keV range and we quote both absorbed and
unabsorbed fluxes. 

\begin{table}[!h]
\caption{Spectral fits to extended sources}
\begin{tabular}{lcccccc}
Object & N$_H$ & $\Gamma$& abs. flux& unabs. flux & $\chi^2$/dof \\
& $10^{21}$cm$^{-2}$& & $f_{0.5-8}^\dagger$&$f_{0.5-8}^\dagger$& \\
\hline
Nebula & $5.0\pm0.4^\ddag$ & $1.77_{-0.08}^{+0.09}$ & $5.6\pm0.23$ & $7.9\pm0.32$ & 67.4/77 \\
Equatorial PWN& $5.0\pm0.4^\ddag$ & $1.57\pm0.08$ & $2.4\pm0.07$ & $3.2\pm0.09$ & 70.3/81 \\
Torus & $5.0\pm0.4^\ddag$ & $1.48\pm0.08$ & $1.5\pm0.05$ & $1.9\pm0.07$ & 39.8/53 \\
\hline
Jet & $5.0^*$ & $1.26_{-0.13}^{+0.14}$ & $0.42\pm0.03$ & $0.52\pm0.04$ & 12.0/18 \\
Counter Jet & $5.0^*$ & $1.39\pm0.10$ & $0.81\pm0.04$ & $1.0\pm0.05$ & 10.8/18 \\
Outer Jets & $5.0^*$ & $1.26\pm0.18$ & $0.27\pm0.03$ & $0.33\pm0.03$ & 11.9/18 \\
\end{tabular}

$^\dagger$ 0.5-8\,keV fluxes in units of $10^{-13}{\rm erg/cm^2/s}$ \qquad $^*$ held fixed
\qquad $^\ddag$ Simultaneous $N_H$ fit.
\label{pwnspec}
\end{table}

	Note that there is significant softening of the extended emission as
one progresses to larger scales and that the jet components appear to be the 
hardest of all. This is certainly consistent with the idea that the central
pulsar supplies fresh energetic electrons and that synchrotron burn-off
increasingly softens the spectrum as older populations are viewed in the outer PWN. 
Again this trend is common in the well-measured young PWNe. Allowing
the photon index to vary for the different nebula components, the best fit 
to a global absorption value for the extended emission gives us our fiducial 
$N_H = 5\times 10^{21} {\rm cm^{-2}}$. This is consistent with free-fit values
for the point source, but given complexities of the composite thermal+power law
model, we consider the nebular fit value more robust. Note that with DM=75.7
${\rm cm^{-3}pc}$, the $H/n_e \approx 21$ for this sight-line is large,
but not unprecedented for low $|b|$ pulsars. This is also consistent with
the  $H_I$ absorption measurements and a fiducial SNR distance $\sim 3$\,kpc, 
given the appreciable uncertainties.

\section{Interpretation and Conclusions}

	A number of authors have discussed the evolution of a PWN within an
expanding supernova remnant. For example, \citet{swa01} and \citet{che05} describe
the early evolution when the supernova ejecta are in free expansion. Later, after
the remnant interior is heated by the passage of the reverse shock, the PWN evolves
within the Sedov phase supernova remnant whose radius is $R_{\rm SNR} = 1.17 
(E_0/\rho)^{1/5} t^{2/5}$ for an explosion energy $E_0$ in a $\gamma=5/3$ medium
of density $\rho$. PSR B1706$-$44 has a characteristic age $10^4\tau_4$\,yr with 
$\tau_4 \approx 1.7$, so G343.1$-$2.3 should be safely in the Sedov phase with
an expected angular size
$$
\theta_{\rm SNR} \approx 16^\prime (E_{51}/n_0)^{1/5} t_4^{2/5}/d_3
\eqno(1)
$$
for a supernova releasing energy $E_0=10^{51}E_{51}{\rm erg}$ in an external 
medium density $n_0\,{\rm cm^{-3}}$, at a true age $10^4t_4$y at a 
distance $3d_3$\,kpc. The observed size then implies 
$E_{51} \approx 11 n_0 t_4^{-2}d_3^5$, requiring a fairly energetic explosion
for $d>2$\,kpc. During the Sedov phase the interior pressure is
$$
P_{SNR} \approx 10^{-9} E_{51}^{2/5} n_0^{3/5} t_4^{-6/5} {\rm g/cm/s^2}
\eqno(2)
$$
and is relatively constant away from the SNR limb.

	The pulsar blows a wind bubble within this SNR interior, whose radius is 
$R_{\rm PWN} \approx (E_\ast/E_0)^{1/3} R_{\rm SNR}$ for a PWN bubble energy
$E_\ast = f {\dot E} \tau_c$ \citep{vdsw01}. Although the accuracy of this dependence
of PWN radius on pulsar injection energy has been questioned \citep{bcf01},
we adopt it for the following estimates.  With the observed ratio of radii,
$R_{\rm PWN}/R_{\rm SNR} = 1.8^\prime/25^\prime$, we obtain $E_\ast = 3.7 \times 10^{-4}
E_0$, i.e. this PWN has quite low internal energy. This is also reflected in the
low radio and X-ray fluxes. Together we use these estimates, the observed
size of the SNR, equation (1) and the measured ${\dot E}_{36}=3.4$ and
$\tau_c= 1.75 \times 10^4$y to write $f \approx (R_{\rm PWN}/R_{\rm SNR})^3
E_0/( {\dot E} \tau_c) = 2.1 \, n_0 t_4^{-2} d_3^5$. Now, if the PWN is adiabatic
and we assume spindown with constant $B$ and braking index $n=3$ from an
initial period $P_0$, we find that the total energy in the plerion is
$[(P/P_0)^2-1] {\dot E} \tau_c$. Then, setting $f=(P/P_0)^2-1$ and eliminating the
true age $t$ using $t=\tau_c[1-(P_0/P)^2]$ for magnetic dipole spindown, we
obtain a constraint on the initial spin period
$$
[1-(P_0/P)^2]^3/(P_0/P)^2 = 0.68\, n_0 d_3^5
\eqno(3)
$$
which has a solution of $P_0=0.61P=62$\,ms for d=3\,kpc, and  $P_0=0.79P=80$\,ms
for d=2\,kpc.  The corresponding true ages are 0.61$\tau_c$ ($1.1\times 10^4$y) 
and 0.38$\tau_c$ ($0.67\times 10^4$y), respectively. These numerical values
are for $n_0=1$ and the density dependence from Equation (3) is quite weak.
\citet{vdsw01} present a similar 
sum for $P_0$, assuming a known $E_0$; the above formulation emphasizes the sensitivity 
to the poorly known $d$. Note that with the large implied initial period, the integrated
PWN energy is quite comparable to the present spin energy, with $f \approx 1.7$ at
d=3\,kpc and $f \approx 0.62$ at d=2\,kpc. So the spindown luminosity is roughly
constant in the adiabatic phase and the PWN growth is closer to $t^{11/15}$ 
than to the $t^{3/10}$ law appropriate for impulsive energy injection \citep{swa01}.

	Inside this wind bubble, the Sedov interior pressure confines the PWN,
giving rise to a termination shock at
$$
\theta_{\rm WS} \approx ( {\dot E} /4\pi c P_{\rm SNR})^{1/2}/d.
\eqno (4)
$$
which results in $\theta_{\rm WS} \approx 1\farcs2 {\dot E}_{36}^{1/2} E_{51}^{-1/5} 
n_0^{-3/10} t_4^{3/5} d_3^{-1}$. If we apply the SNR estimate for $E_0$ above,
this becomes $\theta_{\rm WS}\approx0\farcs72{\dot E}_{36}^{1/2}n_0^{-1/2}t_4 d_3^{-2}$.
Then, using ${\dot E}_{36}\approx 4$ and applying the age estimate following Equation (3) 
we get $\theta_{\rm WS}\approx1\farcs5 n_0^{-1/2}$ (d=3\,kpc) or
$\theta_{\rm WS}\approx2\farcs1 n_0^{-1/2}$ (d=2\,kpc). These estimates are reasonably
consistent with the observed $3^{\prime\prime}$ torus radius, especially since
an equatorially concentrated flow should have a stand-off distance 1.5-2$\times$ this
spherical scale. The polar jets can have an initial shock at somewhat larger
angle, with the resulting pitch angle scattering illuminating the jets somewhat further
from the pulsar.

	Of course, this bubble is offset from the center of G343.1$-$2.3 at 
$R = 0.5\,R_{SNR}$ (figure 1). This is inside the $\sim 0.68\,R_{SNR}$ where \citet{vdsdk04}
note that the increasing density causes the pulsar to be supersonic, so a bow shock
should not
have yet formed. These authors however compute numerical models of a fast moving pulsar
in a SNR interior. As the pulsar moves, the PWN should become highly asymmetric 
with a `relic PWN' 
at the SNR center and the pulsar placed near the leading edge of the PWN; see \citet{vdsdk04} 
figures 7 and 8. We see no PWN structure near the geometric center of G343.1$-$2.3 and, 
if the `bubble nebula' is identified with the shocked pulsar wind, the pulsar is certainly 
not offset from its center along the proper motion axis (away from the SNR center). 
So these models are an inadequate description of G343.1$-$2.3. From Figure 2, the pulsar
is well centered in the bubble nebula, with any offset from its center
along the axis to the SNR substantially less than $30\Delta_{30}$\,arcsec.  Thus
$$
v < 40 \Delta_{30} d_3/t_4 {~\rm km/s},
\eqno (5)
$$
and the pulsar cannot have moved far from the explosion center. This is, of course,
consistent with the scintillation results.

	We can reconcile the symmetric PWN with the offset SNR shell if we assume
that the pulsar progenitor exploded toward the edge of a quasi-spherical
cavity. One scenario (also posited by Gvaramadze 2002, Bock \& Gvaramadze 2002) that can
associate the low velocity pulsar with G343.1$-$2.3 is to assume that the progenitor star 
had a stellar wind of mass loss rate ${\dot M}_{-8}10^{-8}M_\odot$/yr and
wind speed $10^8v_{\rm w8}$\,cm/s over $t \sim t_7 10^7$y, typical of the $\sim 10 M_\odot$
stars that dominate the pulsar progenitors \citep{maed81}. This
evacuates a stellar wind bubble of size
$$
\theta_{SW} = 46^{\prime} \left ( {\dot M}_{-8}v_{\rm w8}^2/n_0 \right )^{1/5} t_7^{3/5}/d_3.
\eqno (6)
$$
During the main sequence lifetime, the star moving at $10v_6$\,km/s
travels $\sim 2^\circ v_6 t_7/d_3$
and so it can easily traverse its wind bubble. Thus, one can imagine an
off-center supernova in a nearly symmetric stellar wind bubble of radius
$\sim 25^\prime$: the supernova blast wave expands to fill the bubble,
passing to the Sedov phase near its present radius. The supernova
produces a neutron star with little or no kick, placing the pulsar near
its present position. This has the added advantage of accommodating the rather
large SNR size with a more modest energy of a few$\times 10^{51}$\,erg. The
PWN energy and size estimates above would then be somewhat amended; this would require a
careful numerical simulation.

	For the reasons detailed in the introduction, it is not yet clear that
PSR B1706$-$44 and G343.1$-$2.3 are associated. So for completeness we can consider
the case when the shocked pulsar wind blows an adiabatic bubble in a static, 
low $P_{\rm ext}$ external medium \citep{cmw75}. If we assume that the pulsar was
born (sans SNR) or entered a confining region of the ISM $\sim 10^4$y ago
and that since then it has been spinning down at the present energy loss rate, we
find that it will blow a bubble of angular size
$$
\theta_{\rm BN} \approx 0.76 ({\dot E}t^3/\rho)^{1/5}/d \approx 
120^{\prime\prime} ({\dot E}_{36}/n_0)^{1/5} \tau_4^{3/5}/d_3.
\eqno (7)
$$
These estimates change somewhat for a pulsar born at $P_0 \ll P$; since we
are not making the association with the SNR G343.1$-$2.3, we can make no 
estimate of the initial spin period.
As first noted by \citet{dg02}, the $\sim 4^\prime$ wide radio spur across the
face of G343.1$-$2.3 has the approximate scale of such a `bubble nebula'. If the
PWN stays unmixed (relativistic) then the interior of the bubble will have
a pressure 
$P_{\rm BN} \approx {\dot E}t/(4\pi R^3) \approx 1.6\times 10^{-10}
\left (n_0^3{\dot E}_{36}^2/\tau_4^4\right )^{1/5} {\rm g/cm/s^2}$. In turn,
the torus termination shock in this medium is at
$$
\theta_{\rm WS} \approx 2\farcs9 ({\dot E}_{36}/n_0)^{3/10} \tau_4^{2/5}/d_3.
\eqno (8)
$$
If (e.g. through Rayleigh-Taylor instabilities) the pulsar wind is well mixed
with the swept up gas, the adiabatic thermal pressure would be $\sim 2\times$ larger.
Interestingly, the angular scales for this scenario are also reasonably compatible
with the observed torus and bubble nebula size. Of course this scenario leaves open
the question of the pulsar origin.  Again the pulsar would need to
have a quite low velocity to produced the observed symmetry.

	Turning to the spectral results, we note that
\citet{poet02} fit a correlation between spindown energy and the 
PSR+PWN luminosity: $L(2-10\,{\rm keV}) =1.8 \times 10^{38}
{\dot E}_{40}^{1.34}{\rm erg/s}$. For the PSR B1706$-$44 parameters this 
predicts a flux $f(2-10\,{\rm keV}) = 4.7 \times 10^{-12}d_3^{-2}{\rm erg/cm^2/s}$. The 
observed 2-10\,keV flux is in fact $\sim 1.7 \times 10^{-12}
{\rm erg/cm^2/s}$, even including the outer `bubble nebula'; without this 
component it is half as large. These correlations are not very accurate,
but this does imply that the PSR B1706$-$44 PWN is substantially under-luminous
for any distance less than 3\,kpc. \citet{got03} has derived
correlations between the pulsar spindown power and the pulsar/PWN spectral indices. 
His relation predicts
$\Gamma_{\rm PSR} = 0.63\pm0.17$ (substantially smaller than our power law index 
$\sim 1.6\pm 0.2$) and $\Gamma_{\rm PWN} = 1.3\pm 0.3$ (not inconsistent with
the values measured for the torus and equatorial PWN).  

As described in \S2.4, the
spectrum softens appreciably from the central torus to the outer bubble nebula.
This suggests increased aging of the synchrotron population. Figures 2 and 3
show that the bulk of the radio emission lies in the `bubble nebula' region.
So we can take the radio flux and spectral index from \citet{giet01}
and compare with our nebula X-ray flux (Fig. 4). Comparing the radio spectral index
$\alpha_R = 0.3$  with the best fit X-ray index $\alpha_X = 0.77$, shows
a break quite close to the $\Delta \alpha=0.5$ expected from synchrotron cooling.
The extrapolated intersection of these power laws gives a break frequency
of Log[$\nu_B {\rm (Hz)}]=12.2^{+0.9}_{-1.1}$. For the fiducial pulsar age
of $\sim 1.7 \times 10^4$y, this corresponds to a nebula field
of $1.4_{-0.6}^{+2.1} \times 10^{-4}$\,G. Note that the magnetic pressure
from this (photon flux-weighted) average field is $\sim 8 \times 10^{-10}
{\rm g/cm/s^2}$, somewhat larger than the nebula pressure estimated from its
radius. This may indicate field compression in the nebula limb.
In general, if the mean nebula field is
$10^{-4}B_{-4}$\,G for a nebula of angular radius $\sim 100^{\prime
\prime}\theta_{100}$, the total nebula field energy is 
$E_B \approx 1.5 \times 10^{47} B_{-4}^2 (\theta_{100}d_3)^3$\,erg.
This is comfortably less than the present spin energy $E_{\rm PSR}\approx 
2 \times 10^{48}$\,erg, so the nebula can be easily powered
even if the pulsar was born close to its present spin period.
We find that this cooling break field is substantially larger than the
equipartition field of $10-15\mu$G inferred for the radio and X-ray
emitting populations (also the minimum equipartition nebula energy
$\sim 9 \times 10^{45}$\,erg is substantially smaller). The cooling
break field can also be compared to that expected from simple
radial evolution of the pulsar surface field: if this field
$B_\ast = 3 \times 10^{12}$\,G falls off as $r^{-3}$ to the light 
cylinder, then as $1/r$ to the wind shock where it is compressed 
we get $B_{\rm WS} \sim 3B_\ast r_\ast^3/(r_{\rm LC}^2 r_{\rm WS})\sim 1$mG.
If it continues to fall off as $1/r$ beyond this we get a field at the
limb of the bubble nebula of $\sim 30\mu$G. 
So the best we can do is to infer a mean nebular field 
$\sim 10-30\times$ the equipartition value, with some generation of
new field beyond the torus wind shock. The energetic requirements for
this field, required to match the $\nu_B$ cooling break, are 
comfortably less than the energy available from PSR B1706$-$44.
These field estimates are consistent with the non-detection of
TeV ICS flux from this source \citep{hess05}.

\begin{figure}[b!]
\includegraphics[scale=0.5,origin=200]{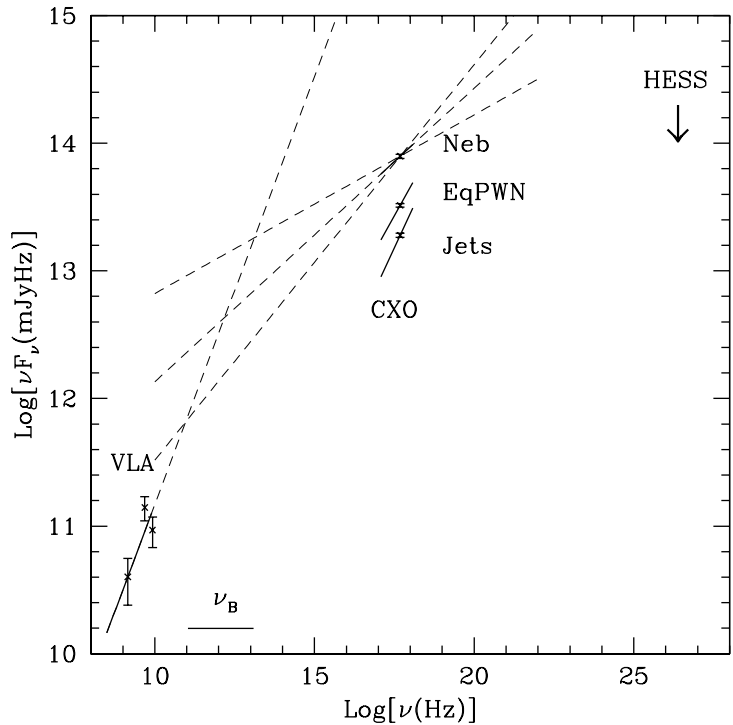}
\caption{Outer PWN (`bubble nebula') spectral energy distribution (SED). 
Radio data are from \citet{giet01}, 
the TeV upper limit is from \citet{hess05}. The spectra of the inner,
younger PWN components are plotted for comparison.
The extension of the radio PL and the best fit nebula (outer PWN)
PL meet at $\nu_B \sim 1.6 \times 10^{12}$\, Hz; the indices are consistent 
with a $\Delta \alpha = 0.5$ cooling break.
}
\end{figure}
\bigskip

	The narrow outer jets also have a power-law spectrum and are almost 
certainly synchrotron-emitting.  For a reasonable $0.1\,B_{-4}$mG field, the
observed X-rays of $E_\gamma = 1.5 B_{-4}\Gamma_{7.5}^2$\,keV require substantial 
$e^\pm$ energies, with $\Gamma_e = 3 \times 10^7\Gamma_{7.5}$
near the radiation-reaction limited primary Lorentz factor inferred for
many polar cap \citep{mh03} and outer magnetosphere \citep{r96} pulsar models.
Since the jet is narrow, confinement of these pairs imposes a (not
very restrictive) lower bound on the jet field
$B_{-4} > 0.075 E_5^{1/3}/(d_3\theta_w)$
where the maximum observed jet photon has $E\sim 5E_5$\,keV and the
observed outer jet half-width is $\theta_w$arcsec. A more restrictive upper
limit on the mean jet field comes from the observation that the jets
do not soften noticeably before their end $\sim 30\theta_{30}$arcsec from
the pulsar. If we assume a jet bulk speed $\beta$c, then arguing that the
flow time is shorter than the synchrotron cooling time gives us the limit
$
B_{-4} < 8.5 \left ( {{\beta} \over {d_3\theta_{30}}} \right )^{2/3} E_5^{-1/3}.
$

	When the observed jet spectrum has an energy index of 
$\alpha = \Gamma -1 \approx 0.3$, we infer a power-law spectrum of $e^\pm$ in the jet
$N(\Gamma_e)d\Gamma_e = K\Gamma_e^{-p}d\Gamma_e$, with $p=2 \alpha +1 \approx 1.6$. We
can then make an estimate of the minimum jet luminosity, i.e. at 
`equipartition' when $B^2 = 6\pi m_e c^2 \int \Gamma_e N(\Gamma_e)d\Gamma_e$.
Given the observed combined outer jet luminosity (0.5-8\,keV)
$L=4\pi d^2f_{\rm oj} \approx 2.7 \times 10^{31} {\rm erg/s}$ and emitting volume
$V \approx 2 \times \theta_L\times \pi\theta_wd^3 \approx 1.1\times 10^{52}
(\theta_L/20)\theta_w^2d_3^3\,{\rm cm^3}$ 
with the angles in arcsec, we can estimate
the equipartition field for an isotropic plasma as
$$
B_{eq} = \left [ {{18\pi} \over {\sigma_T}} 
\left ( {2\pi m_e c}\over {E_{\rm max}}\right )^{1/2}
{{2(1-\alpha)} \over {1-2\alpha}}
{{1-(E_{\rm min}/E_{\rm max})^{(1-2\alpha)/2}}\over{1-(E_{\rm min}/E_{\rm max})^{(1-\alpha)}} }
{L \over V}
\right ]^{2/7}
\eqno (9)
$$
where the observed photon spectrum runs from $E_{\rm min}$ to $E_{\rm max}$.
For the observed flux this gives
$$
B_{\rm eq} \approx 0.25 \times 10^{-4} q(\alpha) [\theta_L\theta_w^2d_3/20]^{-2/7}\, {\rm G},
\eqno (10)
$$
where $q(\alpha=0.3) = 1$ is a weak function of $\alpha$.
The corresponding minimum energy flux for the outer jet is
$$
L_{\rm oj} = 8 \times 10^{33} \beta d_3^{12/7}\theta_w^{10/7}(\theta_L/20)^{-2/7}
{\rm erg/s},
\eqno (11)
$$
where the jet bulk velocity is $\beta$c.
This is  $\sim 10^{-3}{\dot E}$ per jet and will, of course, be larger
if the jet flow includes ions. Interestingly, if the pulsar
couples roughly isotropically to the PWN, then the corresponding fraction
of the outflow should subtend a half angle of $\sim 5^\circ$. This is 
somewhat smaller than the angle subtended by the inner jets, but
$\sim 3\times$ larger than the $\sim 1^{\prime\prime}$ width of the
ends of the jet -- there is substantial collimation of the jet energy flux.

	We have argued that a low PSR velocity can explain the symmetry of the PWN.
The central location of the pulsar and spherical post-shock flow may also allow
the equatorial toroidal structure and polar jets to propagate undisturbed to large radii.
We do, however, measure a small misalignment of the outer jets, corresponding
to a deflection of $\theta_{\rm de} = 1.3\pm 0.15^\circ$ for each. If we imagine
a pressure acting along the jet's $\sim 30^{\prime\prime}$ length, then the
required perturbation is $\delta P \approx L_{\rm oj}{\rm tan}\theta_{\rm de} /
(\beta c A_{\rm oj}) \approx 5 \times 10^{-14} L_{34}/(\beta\theta_{30}\theta_w d_3^2)
{\rm g/cm/s^2}$, where $A_{\rm oj}$ is the jet's cross sectional area.
This is only $\sim 10^{-3}$ of the total pressure in the nebula. It could be
due to ram pressure if the shocked nebular medium flows to the west
at $v \sim 1.7 n_{\rm neb}^{-1/2}$\,km/s. 
\bigskip

	Our X-ray measurements have established the PWN symmetry axis, presumably
reflecting the pulsar spin axis, to very high precision. Unfortunately our
original goal of relating this to the proper motion axis remains unfulfilled.
It is true that the torus symmetry axis points roughly toward the center
of G343.1$-$2.3, confirming the estimates from earlier {\it Chandra} data.
However, the PWN symmetry about the pulsar and the low scintillation velocity suggest
a very low transverse speed $\le 40$km/s, which would preclude a birth site
as distant as the SNR center. This low speed makes a direct proper motion challenging,
but allows latitudinal asymmetries in the PWN flow to propagate undisturbed to
fairly large radius, where they can be imaged with {\it Chandra}. Thus, study
of this PWN offers some good opportunities to probe outflow dynamics and jet
collimation. Study of this, and similar, PWNe may prove useful electrodynamic
analogs of the $10^6\times$ more powerful AGN jets.  A viable scenario for maintaining
the G343.1$-$2.3/PSR B1706$-$44 association posits a supernova event near the 
present pulsar site, with the remnant inflating a pre-existing
off-center cavity. However, the residual (small) proper motion could then have any 
direction. In fact, the faint extension of the PWN (bubble nebula) to the west
and the increased radio surface brightness to the east might suggest a rather slow 
pulsar motion at PA $\sim 80^\circ$. This would be nearly orthogonal to the
torus symmetry axis. With the large $P_0$ estimated here, this could be construed
as suggesting poor rotational averaging of a birth kick \citep{nr04}. So
the PWN/SNR geometry offers both aligned and orthogonal axes.
Only a sensitive astrometric campaign can detect or limit the pulsar motion
and resolve this ambiguity.

\acknowledgments

This work was supported in part by NASA grants SAO G04-5060X and NAG5-13344. We thank the
referee for several careful readings, which resulted in substantial changes. We also
thank R.N. Manchester who provided a current pulsar ephemeris for the VLBI experiment.


\begin{thebibliography}{}

\bibitem[Aharonian et al.(2005)]{hess05}Aharonian, F. et al 2005, A\&A, 432, L9

\bibitem[Bock & Gvaramadze(2002)]{bg02}Bock, D.C.-J. \& Gvaramadze, V.V. 2002, A\&A, 394, 533

\bibitem[Blondin, Chevalier \& Frierson(2001)]{bcf01}Blondin, J.M., Chevalier, R.A.
\& Frierson, D.M. 2001, ApJ, 563, 806

\bibitem[Castor, McCray \& Weaver(1975)]{cmw75}Castor, J., McCray, R. \& Weaver, R. 1975, ApJ,
 200, L107

\bibitem[Chadwick et al.(1997)]{chet97}Chadwick, P.M. et al 1997, in Proc. 26th Cos Ray Conf, 3, 189

\bibitem[Chevalier(2005)]{che05}Chevalier, R.A. 2005, ApJ, 619, 839


\bibitem[Cordes \& Lazio(2002)]{cl02}Cordes, J.M. \& Lazio, T.J.W. 2002, astro-ph/0207156

\bibitem[Dodson \& Golap(2002)]{dg02}Dodson, R. \& Golap, K. 2002, MNRAS, 334, L1

\bibitem[Dodson et al.(2003)]{doet03}Dodson, R., Lewis, D., McConnell, D.  \& 
Deshpande, A.A. 2003, MNRAS, 343, 116





\bibitem[Frater, Brooks \& Whiteoak(1992)]{atca92}Frater, R.H., Brooks, J.W., \& Whiteoak, J.B. 1992, JEEE, 12, 103

\bibitem[Giacani et al.(2001)]{giet01}Giacani, E.B., Frail, D.A., Goss, W.M. \&
Vietes, M. 2001, AJ, 121, 3133

\bibitem[Gotthelf (2003)]{got03}Gotthelf, E.V. 2003, ApJ, 591, 361

\bibitem[Gotthelf, Halpern, \& Dodson(2002)]{ghd02}Gotthelf, E., Halpern, J.P. \& Dodson, R. 2002, ApJL, 567, L125

\bibitem[Gvaramadze (2002)]{g02}Gvaramadze, V.V. 2002, in Neutron Stars in Supernova Remnants, ed. P. O. Slane, \& B. M. Gaensler (San Francisco: ASP), ASP Conf. Ser., 271, 23  




\bibitem[Johnston et al.(1992)]{jet92}Johnston, S. et al 1992, MNRAS, 255, 401

\bibitem[Johnston, Nicastro \& Koribalski(1998)]{jnk98}Johnston, S.,
Nicastro, L., Koribalski, B. 1998, MNRAS, 297, 108

\bibitem[Kifune et al.(1995)]{ket95}Kifune, T. et al. 1995, ApJ, 438, L91

\bibitem[Koribalski et al.(1995)]{koet95}Koribalski, B., Johnston, S., Wiesberg, J.,
Wilson, W. 1995, ApJ, 441, 756

\bibitem[Lai, Chernoff \& Cordes(2001)]{lcc01}Lai, D., Chernoff, D.F. \& Cordes, J.M. 2001, ApJ, 549, 1111

\bibitem[Maeder(1981)]{maed81}Maeder, A. 1981, AA, 102, 401

\bibitem[McAdam, Osborne \& Parkinson(1993)]{mo93} McAdam, W., Osborne, J
\& Parkinson, M. 1993, {\it Nature}, 361, 516
%
\bibitem[McGowan et al.(2004)]{met04}McGowan, K.E. et al 2004, ApJ, 600, 343

\bibitem[Mori et al.(2001)]{mo01}Mori, K. et al 2001, ASPC, 251, 576

\bibitem[Muslimov \& Harding(2003)]{mh03}Muslimov, A. \& Harding, A.K. 2003,
\apj, 588, 430

\bibitem[Ng \& Romani(2004)]{nr04}Ng, C.-Y. \& Romani, R.W. 2004, ApJ, 601, 479

\bibitem[Nicastro, Johnston \& Koribalski(1996)]{net96}Nicastro, L., Johnston, S. \& Koribalski, B. 1996, AA, 306, L49



\bibitem[Pavlov et al.(2003)]{pet03}Pavlov, G.G., Teter, M.A., Kargaltsev, O. 
\& Sanwal, D., 2003, ApJ, 591, 1157

\bibitem[Possenti et al.(2002)]{poet02}Possenti, A. 2003, AA, 387, 993


\bibitem[Rees \& Gunn(1974)]{rg74}Rees, M.J. \& Gunn. J.E. 1974, MNRAS, 167, 1

\bibitem[Romani \& Yadigaroglu(1995)]{ry95}Romani, R.W. \& Yadigaroglu, I.-A. 1995, ApJ, 438, 314.

\bibitem[Romani(1996)]{r96}Romani, R.W. 1996, \apj, 470, 469

\bibitem[Romani(2004)]{r04}Romani, R.W. 2004, Proc Ast. Soc. Pac., 328; astro-ph/0404100



\bibitem[Spruit \& Phinney(1998)]{sp98}Spruit, H. \& Phinney, E.S. 1998, {\it Nature}, 393, 139

\bibitem[van der Swaluw(2001)]{swa01}van der Swaluw, E., Achterberg, A., Gallant, Y.A., Toth, G. 2001 A\&A, 380, 309 

\bibitem[van der Swaluw, Downes \& Keegan(2004)]{vdsdk04}van der Swaluw, E., Downes, T.P
\& Keegan, R. 2004, A\&A, 420 937

\bibitem[van der Swaluw \& Wu(2001)]{vdsw01}van der Swaluw, E. \& Wu, Y. 2001, ApJ, 555, L49

\bibitem[Zavlin et al.(1996)]{zet96}Zavlin, V.E. et al 1996, AA, 315, 141

\end{thebibliography}
\end{document}